\documentclass{aa}  
\usepackage{graphicx}
\usepackage{natbib}
\usepackage{longtable,lscape}
\usepackage{rotating}

\usepackage{txfonts}
\newcommand{\Msun}{\mbox{$\rm M_\odot\,$}}

\newcommand{\Rsun}{\mbox{$\rm R_\odot\,$}}

\def\arcsec{\hbox{$^{\prime\prime}$}}

\begin{document}
\title{Spitzer-IRAC GLIMPSE of high mass protostellar objects. II} 
\subtitle{SED modelling of a bonafide sample}
   \author{J. M. C. Grave\inst{1,2}
          \and M. S. N. Kumar\inst{1}
		}

   \offprints{J. M. C. Grave; email:jgrave@astro.up.pt}

   \institute{Centro de Astrof\'{i}sica da Universidade do Porto, 
              Rua das Estrelas, 4150-762 Porto, Portugal
	  \and Departamento de Matem\'{a}tica Aplicada da Faculdade de Ci\^{e}ncias da Universidade do Porto, Portugal\\
	  \email{jgrave@astro.up.pt; nanda@astro.up.pt}}
\authorrunning{Grave \& Kumar}
\titlerunning{ SED modelling of massive protostars}
 %  \date{Received September 15, 1996; accepted March 16, 1997}
\date{\today}
% \abstract{}{2}{3}{4}{} 
% 5 {} token are mandatory
 
  \abstract
  { In a previous work (paper I) a sample of 380 HMPO targets was
    studied using the GLIMPSE point source catalog and images.
    Colour-magnitude analysis of the point sources resulted in the
    identification of infrared counterparts (IRC) of the (sub)mm cores
    of HMPO candidates which were considered bonafide targets.}
{We aim to estimate and analyse the physical properties of the
  infrared counterparts of HMPOs by comparing their spectral energy
  distributions (SED) with those predicted by radiative transfer
  accretion models of YSOs.}
{The SED of 68 IRC's are extended beyond the GLIMPSE photometry to the
  possible limits, from the near-infrared to the millimetre
  wavelengths by using the 2MASS, GLIMPSE version 2.0 catalogs, MSX,
  IRAS and some single dish (and interferometric) (sub)mm data. An
  online SED fitting tool that uses 2D radiative transfer accretion
  models of YSOs is employed to fit the observed SED to obtain various
  physical parameters.}
{ The SED of IRC's were fitted by models of massive protostars with a
  range of masses between 5--42 \Msun and ages between $10^3$ and
  $10^6$ years. The median mass and age are 10\Msun and 10$^4$yr's.
   The observed data favours protostars of low effective
    temperatures (4000-1000K) with correspondingly large effective
    photospheres (2-200\Rsun) for the observed luminosities. The
  envelopes are large with a mean size of $\sim$0.2-0.3\,pc and show a
  distribution that is very similar to the distribution of the sizes
  of 8$\mu$m nebulae discussed in Paper I. The estimated envelope
  accretion rates are high with a mean value of 10$^{-3}$\Msun /yr and
  show a power law dependence to mass with an exponent of 2,
  suggesting spherical accretion at those scales.  Disks are found to
  exist in most of the sources with a mean mass of $10^{-1.4\pm0.7}$
  \Msun.}
% conclusions heading (optional), leave it empty
{ The observed infrared-millimetre SED of the infrared counterparts of
  HMPOs are successfully explained with an YSO accretion model. The
  modelled sources mostly represent proto-B stars although some of
  them could become O stars in future. We demonstrate that many of
  these results may represent a realistic picture of massive star
  formation, despite some of the results which may be an effect of the
  assumptions within the models. }


   \maketitle
%
%________________________________________________________________

\section{Introduction}

The {\textit Spitzer Space Telescope} has produced a wealth of data in
the infrared bands with improved spatial resolution, providing new
insights to our understanding of the processes involved in massive
star formation. Using the Galactic Legacy Infrared Mid-Plane Survey
Extraordinaire (GLIMPSE) \citet{benjamin03} data which exclusively
covers the massive star forming content of our galaxy, we started the
analysis of infrared point sources and nebulae associated with
candidate high mass protostellar objects (HMPOs) available in the
GLIMPSE fields \citep{kg07} (hereafter Paper I). Our sample of HMPOs
was obtained by combining four surveys, two from the northern
hemisphere \citep{mol96,sri02} (hereafter Mol96 and Sri02) and two
from the southern hemisphere \citep{fau04,fon05} (hereafter Fau04 and
Fon05), that were selected by far-infrared colour criteria. In Paper
I, using the GLIMPSE point source catalog and images we identified
several infrared point sources and compact nebulae associated with
candidate HMPO fields. A good fraction of point sources were found to
have 3.6--8.0$\mu$m spectral indices and magnitudes representative of
luminous protostars with masses greater than 8 \Msun. The GLIMPSE
images revealed compact nebulae centred on the HMPO targets with
striking morphological resemblances to known classes of UCHII regions,
suggesting that the nebulae are probable precursors to the UCHII
regions.

In this paper, we extend upon that analysis to obtain more quantitative
results by complementing the GLIMPSE data with data in other bands and
by modelling the spectral energy distributions (SED) of the infrared
counterparts of high mass protostellar objects (HMPO IRCs).  The
colour and spectral index analysis presented in Paper I just
represents the general nature of the point sources. By constructing
and analysing a wide SED, it is possible to quantify several physical
parameters and also constrain their evolutionary stage
\citep{fazal08}. Such analysis, however, requires not only a good
coverage of the wavelength range, but also high spatial resolution
data to ensure that the fluxes we are studying arise mainly from the
star-disk-envelope system and are not contaminated by their
surroundings. For this purpose, using the infrared surveys such as
2MASS, GLIMPSE, MSX, IRAS and several mm and sub-mm surveys from the
literature, we assembled the best data available for a sample of
bonafide IR counterparts of HMPO candidates and construct their SED to
the best possible extent.

Recently, there has been significant improvements in radiative
transfer modelling of the SEDs of young stellar objects (YSOs), based
on the physics of star formation that we have learnt in the past
couple of decades. An online SED fitting tool has been successfully
developed and tested on the SED of low mass young stellar objects by
\citet{rob07}. This tool uses a grid of 2D radiative transfer models
of YSOs \citep{rob06} that were developed by \citet{whi03a,whi03b}.
These consist of a central star surrounded by an accreting disk, a
flattened envelope and bipolar cavities with radiative equilibrium
solutions of \citet{bjo01}. Although these models successfully
estimate the physical parameters and consistently explain the SED of
low mass YSOs, the assumed physics is thought to be valid for
protostars of masses up to 50\Msun. This in part is due to the
  observational evidence of massive (50-80\Msun) molecular outflows
  \citep{she00,zha07}, disks and toroids around massive protostellar
  candidates \citep{bel05,bel06} and in part due to the requirement
  from theoretical models \citep{ys02,mt03}.

 In this paper, this fitting tool will be applied to the
SED of our bonafide infrared counterparts (IRCs) of HMPO candidates
and the resulting parameters will be analysed in an attempt to understand the
nature of the HMPO IRCs and also to asses the reliability of the
assumed physics in explaining the high mass protostars. Distinction is
made between the results that are independent of the assumed physics
in the models and the contrary.

In Sec.2 we will describe the selection criteria for ``the bonafide''
list of targets, the data used to build the SED of the HMPO IRC's, and
discuss the aspects of the SED fitting tool pertinent for this
analysis. In Sec.3 the modelling procedure is described and we explain
how to separate the model grid dependent biases. In Sec.4 the results
are presented and in Sec.5 we discuss the caveats of the data and the
methods to identify the unbiased results.

\section{Data Selection}

In paper I, several IRC's were identified based on the spectral
indices computed using the IRAC photometry. An alpha-magnitude
($\alpha$=IRAC bands spectral index) product (AM product) was defined
to select the most embedded and luminous sources. The tight positional
correlation of such sources with the HMPO IRAS and mm sources enabled
us to define them as the IRC's of HMPOs. Using these high AM product
IRC's as the starting point, we now construct SEDs of massive
protostars with the criteria that photometry is available in at least
one of the (sub)mm bands. For this purpose, we have examined each of
our fields using GLIMPSE images and (sub)mm maps and re-identified the
IRCs based on positional correlation constraints over the near
infrared to millimetre range.  In paper I we had used version 1.0 of
the GLIMPSE catalog which was a March 2007 delivery. In this paper, we
use the version 2.0 data release (April 2007, April 2008) which has
resulted in several extra sources that did not appear in paper I. The
new sample also contains sources that were previously not classified
as high AM sources because of their saturated nature and/or
missing/poor photometry in the version 1.0 catalog.

\subsection{Selecting the IRCs}

Photometry in as many possible bands from the near-infrared to the
millimetre range was obtained by using the 2MASS, GLIMPSE I (both
highly reliable and less reliable), MSX, IRAS and the (sub)mm surveys
that are the basis of the HMPO sample.  The mm/sub-mm sources were
found with SCUBA and MAMBO for the Sri02 \citep{wil04,beu02} and the
Mol96 surveys \citep{mol00} and with SIMBA for the Fau04 and the Fon05
\citep{bel06} surveys.  As a result, for each source, we obtained
fluxes with a wavelength coverage between 1.2$\mu$m and 1.2mm, namely
on the 2MASS J, H, K bands, Spitzer 3.6, 4.8, 5.6 and 8.0 $\mu$m
bands, MSX A, C, D and E bands, IRAS 12, 25, 60 and 100 $\mu$m bands
and the mm/sub-mm bands of SCUBA, MAMBO and SIMBA cameras. The 2MASS
and GLIMPSE photometric points are measurements within an aperture of
1.2\arcsec and 2.4\arcsec respectively, and the sub-mm/mm fluxes are
extractions within an aperture $\le$8\arcsec--14\arcsec corresponding
to the 450 $\mu$m, 850 $\mu$m and 1.2mm bands. Two sources namely
IRAS18089 and IRAS19217 have been observed with mm interferometers and
the single dish data for these sources are complemented by
interferometric observations \citep{beu04,beu05}. These
  observations had beam sizes less than 2\arcsec and we set the
  aperture size equal to the GLIMPSE data aperture size of 2.4\arcsec.
  This is because, the SED fitting tool simple requires the measured
  flux to be obtained from an aperture smaller than the one used for
  fitting.  Note that we did not use the exact aperture size because
  that would only overplot an extra fit for that aperture.

The identification of the IRC's to construct their SEDs was made based
on the following rules of thumb. In most cases, we begin by identifying
the closest GLIMPSE source to the (sub)mm peak. After this, we match this
source with the 2MASS, MSX and IRAS point source catalogs to identify
the counterparts in each of these bands and obtain the fluxes. The
matching procedure resulted in the following three categories:
\begin{itemize}

\item{ The most common and unambiguous category where a previously
    classified or newly found high AM source coincided with the
    (sub)mm peak to better than 2\arcsec-3\arcsec and also with the
    MSX and IRAS peaks.}
\item{ The case when the offsets of the GLIMPSE source was larger than
    2\arcsec-3\arcsec.}
\item { Two or three bright and red GLIMPSE sources coincide with
   the (sub)mm peak to a positional accuracy of better than half
   width of the (sub)mm beam.}

\end{itemize}

In the second category where the GLIMPSE point source offset from the
(sub)mm peak was larger than a few arcsec, we made a judgement based
on the colours of the source and the beam size of the (sub)mm data
used. If only a single unambiguous red and luminous object was found
within roughly one beam size, the (sub)mm emission was attributed to
that source and chosen for subsequent modelling. In the case where
even two or more sources were found, showing red colours and causing
ambiguity in associating with the (sub)mm peak, we rejected the case
and did not model any source.

\begin{sidewaysfigure*}
\centering

\includegraphics[width=20.5cm,angle=0]{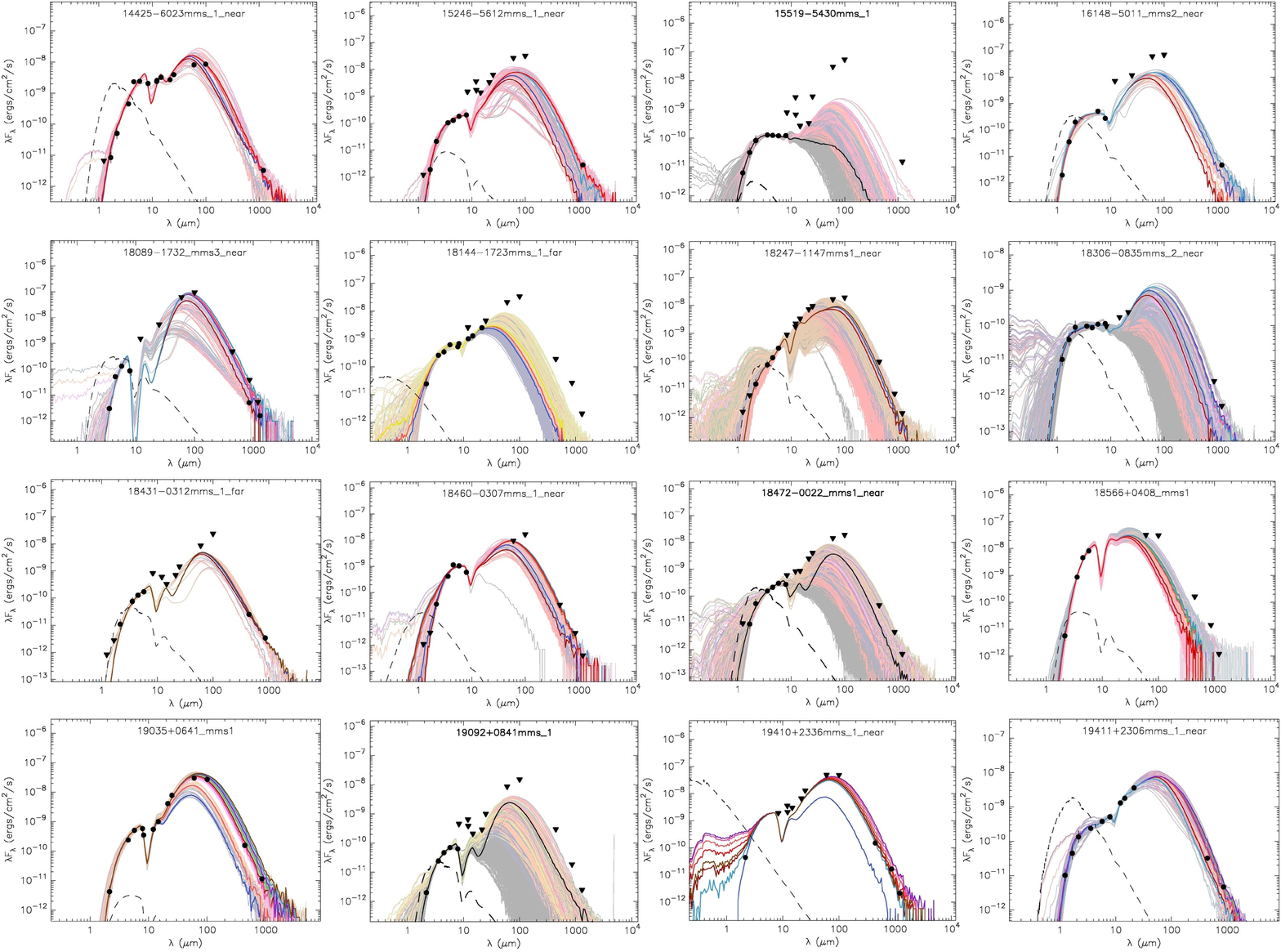}
%\vskip -3cm
\caption{A sample of observed SED and their best fit models. Filled
 circular symbols represent the data points and the filled triangles
 represent photometry used as upper limits on the fitting. The
 dashed line corresponds to the stellar photosphere model used in the
 best YSO model fit. }
     \label{fig:seds}
\end{sidewaysfigure*}

Only a few exceptional sources(e.g. IRAS18454) fell in to third
category where two or more red sources were found very close to the
(sub)mm peak. In such cases, the reddest and brightest GLIMPSE point
sources closest to the peaks were modelled by using the (sub)mm, MSX
and IRAS fluxes as upper limits.

\subsection{Data points and upper limits}

The SED fitting tool developed by \citet{rob07} was used to fit the
data with a grid of YSO models presented by \citet{rob06}. This tool
requires at least three wavelength points to fit the SED and any
number of fluxes that can represent upper limits. The constraint on a
data point is that the photometry extracted within a defined aperture
must by all criteria represent a single source, with measured fluxes
and associated errors.  The good quality GLIMPSE or 2MASS data points,
which have the highest spatial resolution among all the available data
usually were used as the ``data points''.  However, we adopted a
strategy to make the best use of all the data in constraining the SED.
In many cases, we have used the (sub)mm data as a ``data point''
despite their large beam sizes compared to the GLIMPSE or 2MASS
aperture sizes. If the (sub)mm emission indicated a single isolated
clump with a well defined peak and, if within the spatial limits of
the identified clump only one unambiguous high AM product GLIMPSE
source was found, without any contamination from clustering of fainter
objects, then we treated the (sub)mm data as a point source. In those
cases where the (sub)mm emission was extended (despite having a single
peak) and/or if there were more than one red source or fainter group
of stars within the extent of the clump, we treated the (sub)mm data
as upper limits.  Further, we use the peak fluxes of the (sub)mm data
and the beam size as an aperture in contrast to using the integrated
flux from the entire clump.  The MSX and IRAS fluxes were used as upper
limits except for the very few cases where the source appeared totally
unambiguous and isolated satisfying all the selection criteria.  10\%
errors on the fluxes were assumed GLIMPSE data and 50 \% were used for
2MASS fluxes due extinction uncertainties. MSX and IRAS flux errors
are typically around 10\% depending on the bands and quality. The
millimetre observations from the single dish telescopes usually quote
conservative flux errors that were taken from the original literature.
All the fluxes used as upper limits were set with confidence levels of
100\%. These error assumptions provide sufficient margin to account
for poor photometry and/or variability of sources.

\section{SED fitting analysis}

The online SED fitting tool was fed with the data selected according
to the criteria discussed above with appropriate treatment as points
or upper limits.  For each run of fitting, the tool retrieved {\em a
  best fit model} (with the least $\chi^2$ value) and {\em all the
  models} for which the difference between their $\chi^2$ value per
data point and the best $\chi^2$ per data point was smaller than 3.
The number of such models for each target is called the fitting
``degeneracy''.  This is similar to the approach used by \citet{rob07}
in fitting the low mass YSO SEDs.  This approach is taken because the
sampling of the model grid is too sparse to effectively determine the
minima of the $\chi^2$ surface and consequently obtain the confidence
intervals. A total of 68 sources was fitted: 50 from Sri02, 3 from
Mol96, and 15 from Fon05. For all the targets (51) with distance
ambiguity (where a near and far distance estimate was available), the
SED fitting was done independently for the two distance values. The
ambiguity was resolved for the remaining 17 sources. Since a distance
range is used as an input in the fitting tool, we assumed a 20 \%
uncertainty for each of the assumed distances.

Fig.~\ref{fig:seds} shows the results from the SED fitting procedure.
A sample is shown in the printed version and all the figures are
presented online. Circles correspond to the photometric data points
with respective error bars. In most of the cases, the error bars are
so small that they are hidden behind the circles. The triangles are
data points used as upper limits. The dashed curve in each figure show
the stellar photosphere model used in the best fit models.

\begin{figure}

      \includegraphics[width=9cm,angle=0]{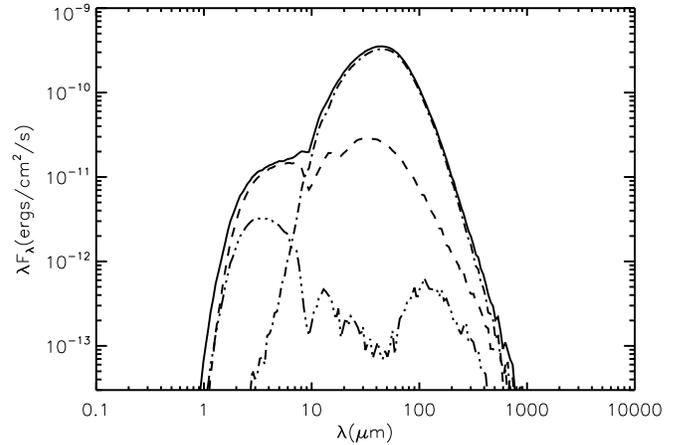}
      \caption{Best fit model for the source 19411+2306mms\_1. The
        full line represents the total flux of the source, the disk
        flux is represented by the dashed line, the envelope flux by
        the dot-dashed line and the scattered flux by the three
        dot-dashed line.}
      \label{fig:sed_example}

\end{figure}

In Fig.\ref{fig:sed_example}, we present the best fit model for the
source 19411+2306mms1. The flux from the main components of the YSO
is shown, namely the disk flux (dashed line), the envelope flux
(dot-dashed line) and the scattered flux (three dot-dashed line).
This example illustrates how different physical components dominate
the emitted radiation at different wavelengths. The disk emission
almost overplots the total flux in the near-IR domain whereas the
envelope emission dominates the mid-IR to the millimetre range.

\begin{figure}
\centering

\includegraphics[width=9cm,angle=0]{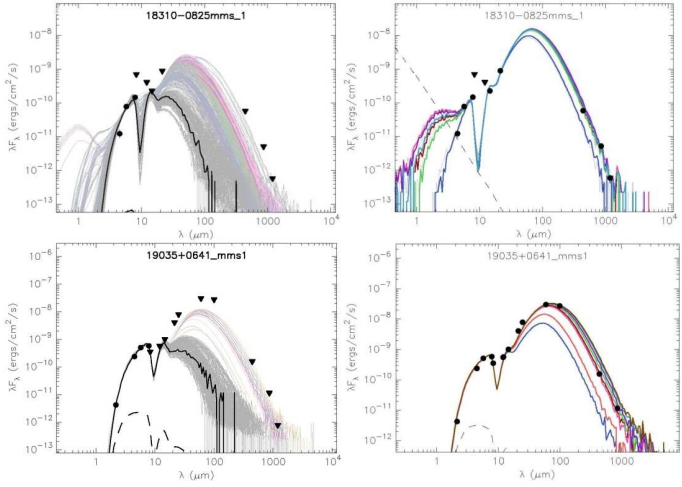}

\caption{Examples of model fitting where the mid, far-IR and mm were
  used as upper limits (left) and data points (right).}

      \label{fig:2seds} 
\end{figure}

Although, the $\chi^2$ criterion is a good representation of the
goodness of fit and associated errors, it is interesting to note that
in a few cases, the model with the lowest $\chi^2$ may not actually
represent the data very well. In some cases, the results were
reiterated because the initial fits showed important information about
the photometry classification. In Fig. \ref{fig:2seds} we show
examples of two sources. The left panels displays the initial fit for
which a low $\chi^2$ value and a high degeneracy was obtained. In
these cases, due to our selection criteria described in Sec.2, the FIR
and (sub)mm data were treated as upper limits. The model with the best
$\chi^2$ (bold line) deviates significantly from the upper limits and
appears not to be a good representation of the observed data. Although
we use certain data as upper limits because of the uncertainties
arising due to lack of information, we know that these may indeed be
used as points if more information was available. In the right
  panel, we force the fit to go through the upper limits which results
  in a slightly higher $\chi^2$ value. Such a forced fit appears to
  better represent the observed fluxes at all wavelengths. For the
  purpose of overall analysis we use these fits rather than the best
  $\chi^2$ fit model.

\subsection{Estimating the physical parameters}

\begin{figure}

      \includegraphics[width=9cm,angle=0]{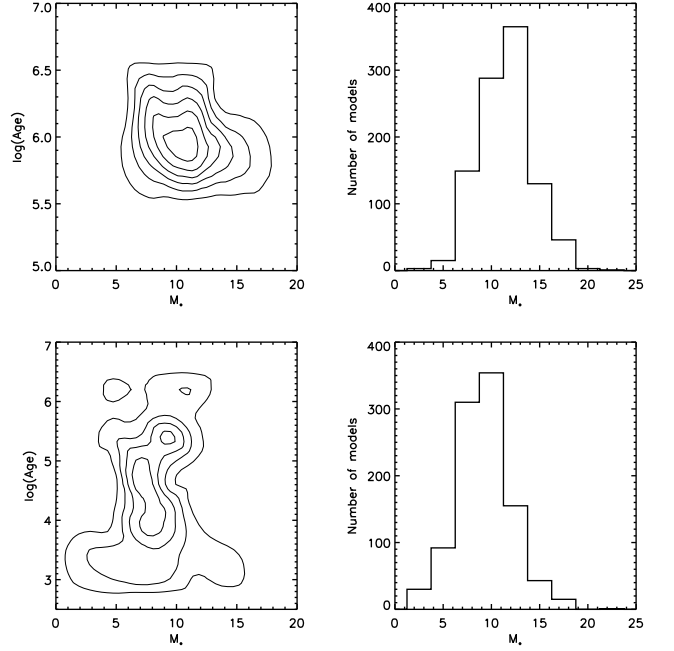}
      
      \caption{Right panels: Histogram of stellar mass estimated from
        all fitted models (within the $\chi^2$ criteria) to the SED of
        sources 16219-4848mms1 (top) and 18290-0924mms1 (bottom). Left
        panels: Correlation of stellar mass vs. age estimated from all
        the fitted models for the source 16219-4848mms1 (top) and
        18290-0924mms1(bottom).}

     \label{fig:corr_deg}
\end{figure}

The observed SED is typically fitted by multiple models, each model
describing a set of physical parameters. The same parameters from
different models can have a wide range spanning several factors to
orders of magnitudes. To identify the representative values of
different physical parameters for each source, the distributions of
each parameter were analysed. As shown in Fig.~\ref{fig:corr_deg}, a
single parameter histogram or a two parameter distribution plot were
used to analyse the their distributions, specifically to see if the
parameters coming from all the models fitting a given source had clear
concentrations or significant spreads. The distributions of the
stellar mass, the age and the luminosity generally show clear peaks
that are independent of the model grid sampling characteristics.
However, parameters such as disk mass or envelope accretion rate do
not show such clear concentrations and usually have skewed
distributions or spread out over a wide range of values.  The right
column in Fig.~\ref{fig:corr_deg} shows the histogram of stellar mass
estimated from all fitted models for sources 16219-4848mms1 (left
panel) and 18290-0924mms1 (right panel). The distribution can be found
to peak at 11 and 9 \Msun for the two sources respectively.  These
distributions are not an effect of the model grid parameter space
coverage but a representative value that best describes the observed
data. In the next section (Sec.3.2), we will demonstrate how we
separate such biases.  In fact the number of models in the grid
decreases constantly with mass (see left plot in figure
\ref{fig:hist}). The left panels in Fig.~\ref{fig:corr_deg} display
the correlation plots of age vs. stellar mass estimated from all the
models that fitted the same two sources. A single concentration can be
seen in this plot for the source 16219-4848mms1(top left) which
represents an age of $10^{5.9}$ years and mass of 11 \Msun. In the
case of 18290-0924mms1 (bottom left panel), the distribution has a
relatively larger spread in the parameter space with a wide
concentration between 6 and 10\Msun and ages between $10^{3.5}$ and
$10^{5.5}$ years. In order to find a representative value for these
distributions, we have decided to compute a weighted mean and standard
deviation for all the parameters of each source, with the weights
being the inverse of the $\chi^2$ of each model. All models where the
${\chi^2}_{best} - \chi^2$ per data point is smaller than 3 are used
to compute the weighted means and standard deviations. It is important
to note that these means are computed on a log scale for quantities
that are spread over orders of magnitude in the model grid, such as
age and accretion rates. This is done because these parameters in the
grid were usually sampled with an approximately constant density in
the log space.

\subsection{What is model grid dependent and what is not?}

The basis of the SED fitting analysis here, is a radiative transfer
model grid, computed assuming certain physics which are thought to be
quite valid for stars with a few solar masses (low mass stars).  The
model grid assumes that the same physics could be valid to understand
the formation of stars up to 50\Msun\. The reason why one expects
that the assumed accretion scenario may work for such high masses is
based on the observational evidence of outflows, disks and toroids
around massive protostellar candidates. Such evidences are now
abundant and has been obtained with spatial resolutions of an
arcsecond or better, tracing size scales as small as a few hundred to
1000AU \citep{she00,zha07}.

The SED fitting tool used here, basically contains a database of
200,000 models, each with different physical properties and the
observed SED is simply compared to the models in the grid to extract
the closest matches. Due to the limitations inherent in the grid of
models, various trends of parameter space are automatically generated
and can be mistaken for a real trend \citep[for more details
see][]{rob08}. It is however, still possible to obtain useful
information from the model grid by comparing the observed trends
against the inherent trends in the grid. For this purpose and
throughout this work, the parameter space from the {\em full grid} is
compared to the derived values of the best fit models.  By making such
a comparison, it will be possible to visualise if the observed trend
or a representative value is due to inherent biases in the model grid
or the contrary. We will see in the next sections that the observed
data indeed preferentially picks up certain narrow constraints on the
physical properties of YSO components, even when the model grid
provides an uniform or wide range of options.

\section{Results}

In Table.\ref{tab:param} (sample shown in printed text), we list the
parameters from the best fit models for the sources shown in Fig.\,1.
The online table contains all the columns originally given by the SED
fitting tool.  The values listed in each column are the weighted means
of the respective parameters as described in Sec.\,3.1.  Additionally,
each column is accompanied by a corresponding extra column quoting the
standard deviation. The various columns in the sample table of the
printed text are as follows: [col. 1] name of the source;[col. 2]
$\chi^2$ per data point; [col.  3] mass of the star; [col.  4] age of
the star; [col.  5] radius of the star; [col.  6] temperature of the
star; [col. 7] envelope accretion rate; [col.  8] disk mass ; [col. 9]
disk accretion rate.

\begin{table*}
 \centering
 \caption{Parameters obtained for 16 sources within our sample. For each parameter we present the weighted average and the weighted standard deviation using all the models which fitted the data of each source within the $\chi^2$ criteria mentioned in the text. When no 'far' or 'near' reference is used in the name of the source, the distance ambiguity is resolved for that source. The values of $\chi^2$ tabulated are per data point.}
  \label{tab:param}
\begin{tabular}{lccccccccccccccc}

  \hline\hline

Name & $\chi^2$ & \multicolumn{2}{c}{Age} & \multicolumn{2}{c}{$M_*$} &  \multicolumn{2}{c}{$R_*$} &  \multicolumn{2}{c}{$T_*$} & \multicolumn{2}{c}{$\dot{M}_{env}$}  & \multicolumn{2}{c}{$M_{disk}$}  & \multicolumn{2}{c}{$\dot{M}_{disk}$} \\ 
&&\multicolumn{2}{c}{log(yr)} & \multicolumn{2}{c}{\Msun} & \multicolumn{2}{c}{log(\Rsun)} & \multicolumn{2}{c}{log($T_{\odot}$)} & \multicolumn{2}{c}{log(\Msun $yr^{-1}$)} & \multicolumn{2}{c}{log(\Msun)} & \multicolumn{2}{c}{log(\Msun $yr^{-1}$)} \\

\hline

14425-6023mms1near & 5.28 & 3.7 & 0.6 & 14 & 3 & 1.9 & 0.4 & 3.7 & 0.2 & -3.1 & 0.4 & -0.9 & 0.7 & -4.5 & 1.0 \\ 
15246-5612mms1near & 0.38 & 4.1 & 0.4 & 11 & 2 & 1.6 & 0.3 & 3.8 & 0.2 & -2.9 & 0.7 & -1.1 & 0.8 & -5.2 & 1.2 \\ 
15519-5430mms1 & 0.03 & 6.2 & 0.4 & 6 & 1 & 0.5 & 0.2 & 4.2 & 0.1 & -6.7 & 1.1 & -2.5 & 1.0 & -7.9 & 1.3 \\ 
16148-5011mms2near & 2.24 & 4.2 & 0.3 & 11 & 1 & 1.7 & 0.2 & 3.8 & 0.1 & -2.6 & 0.3 & -1.4 & 0.8 & -5.5 & 1.2 \\ 
18089-1732mms3near & 4.16 & 3.7 & 0.5 & 12 & 5 & 1.8 & 0.4 & 3.8 & 0.3 & -3.5 & 0.7 & -0.6 & 0.5 & -4.1 & 0.6 \\ 
18144-1723mms1far & 1.41 & 5.2 & 0.4 & 18 & 2 & 0.8 & 0.1 & 4.5 & 0.1 & -4.2 & 0.9 & -1.1 & 0.6 & -5.4 & 0.8 \\ 
18247-1147mms1far & 0.00 & 4.6 & 0.7 & 19 & 4 & 1.1 & 0.6 & 4.3 & 0.4 & -3.6 & 0.4 & -1.6 & 0.8 & -6.0 & 1.1 \\ 
18306-0835mms2near & 0.48 & 6.0 & 0.6 & 7 & 1 & 0.6 & 0.3 & 4.3 & 0.2 & -6.0 & 1.5 & -2.5 & 1.0 & -7.7 & 1.2 \\ 
18431-0312mms1far & 2.69 & 3.9 & 0.6 & 12 & 2 & 1.8 & 0.4 & 3.8 & 0.2 & -2.8 & 0.1 & -0.6 & 0.3 & -4.4 & 0.7 \\ 
18460-0307mms1near & 3.78 & 5.0 & 0.4 & 13 & 3 & 0.8 & 0.2 & 4.4 & 0.1 & -3.4 & 0.3 & -1.0 & 0.6 & -5.5 & 0.8 \\ 
18472-0022mms1near & 0.08 & 5.6 & 0.9 & 8 & 2 & 0.8 & 0.4 & 4.2 & 0.2 & -5.0 & 1.3 & -2.0 & 0.9 & -7.0 & 1.2 \\ 
18566+0408mms1 & 1.12 & 3.7 & 0.4 & 40 & 6 & 1.8 & 0.6 & 4.2 & 0.3 & -3.8 & 0.1 & -0.9 & 0.7 & -5.3 & 0.6 \\ 
19035+0641mms1 & 6.65 & 4.7 & 0.1 & 11 & 0 & 1.2 & 0.1 & 4.1 & 0.1 & -2.6 & 0.0 & -2.0 & 0.3 & -6.0 & 1.0 \\ 
19092+0841mms1 & 0.08 & 6.0 & 0.8 & 9 & 2 & 0.7 & 0.3 & 4.3 & 0.2 & -5.1 & 1.4 & -2.1 & 1.0 & -6.8 & 1.3 \\ 
19410+2336mms1near & 43.87 & 4.8 & 0.0 & 11 & 0 & 1.1 & 0.0 & 4.2 & 0.0 & -2.5 & 0.0 & -1.7 & 0.0 & -5.4 & 0.0 \\ 
19411+2306mms1near & 1.74 & 3.9 & 0.4 & 10 & 1 & 1.8 & 0.2 & 3.7 & 0.1 & -3.3 & 0.3 & -0.8 & 0.6 & -5.1 & 0.9 \\

\hline
\end{tabular} 
\end{table*}

As we mentioned before, the SED fitting was carried out using both the
near and far distance estimates. The overall statistical analysis of
the YSO physical parameters was made by using the fit with a distance
that provided the best $\chi^2$ per data point for each source. The
exception is Fig.~5 and the first two panels of Fig.~7 where we
separate the distributions of the respective parameters in near and
far distance results. (In the online table we also present the the
results for both distances for the sources to which it applies.){\em
  Therefore the general trends that we explain in the rest of this
  paper will have a mixture of near, far and distance resolved
  fittings.}

The model provides estimates of the interstellar extinction, the
extinction interior to the envelope and the luminosity of the source.
The interstellar extinction Av is estimated to span over a range
between 0-150, and the histogram shows a peak between 10 and 25 mags.
The distribution of the extinction values internal to the modelled YSO
is found to peak between 30 and 300 mag. This is roughly similar to
the extinction estimated from the millimetre continuum observations of
such cores.

  In the following we will elaborate on the overall results pertinent
  to the main components of the protostellar SED.

\begin{figure*}
\centering

      \includegraphics[width=18cm,angle=0]{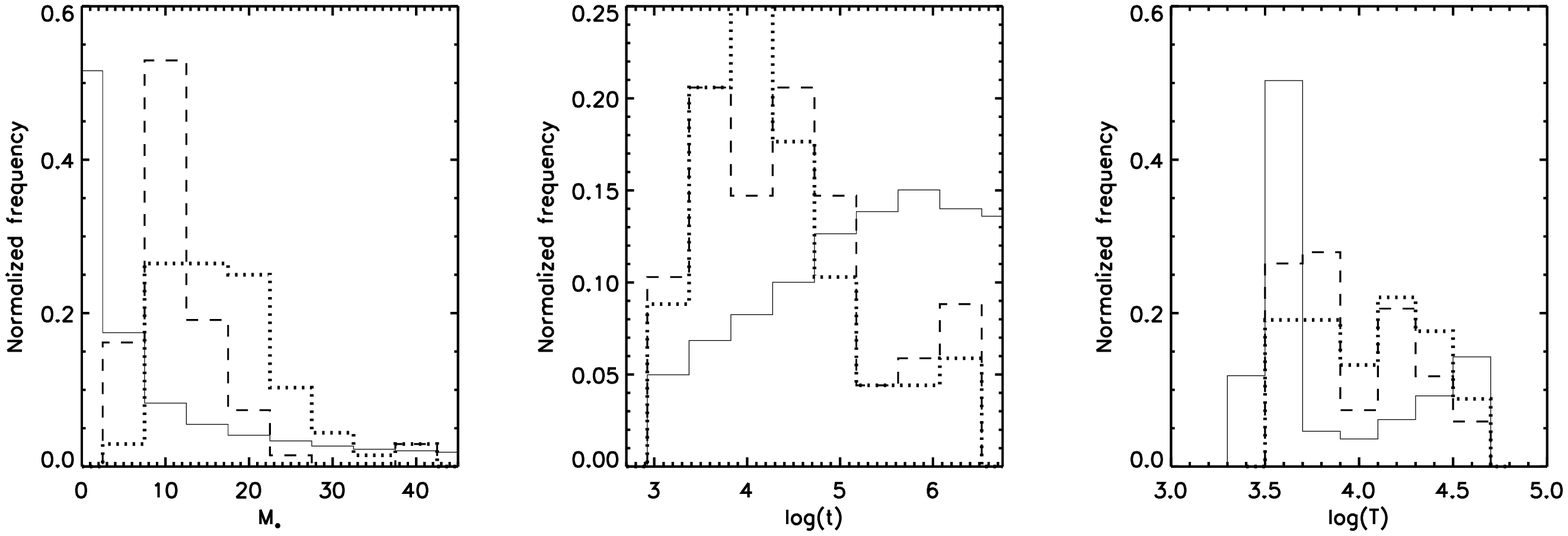}

      \caption{Histograms of stellar mass (left), age (centre) and
        temperature (right) for all the 68 sources in our sample. The dashed histograms in the 3 plots represent the same distributions for all the models in the grid of \citet{rob06}.}

         \label{fig:hist}
   \end{figure*}

\subsection{Properties of Photospheres (driving engines)}

The photospheres are the driving engines of massive protostars which
is also the end product of the massive star formation process that we
seek to understand. The embedded photospheres of the massive
protostellar candidates are already expected to be massive even though
it is accreting material.  Using the weighted mean values of the
parameters for the photosphere from the SED fitting procedure, we can
investigate the general properties of such massive stars.

Fig.~\ref{fig:hist} shows the overall statistics based on the
estimated representative values for the stellar mass, age and
temperature for all the fitted sources.  These histograms are shown
with a normalised frequency in order to compare them with the grid
distribution. The dashed line represents the results from the near
distance fittings, dotted lines for the far distance fittings and the
solid line represents the histograms of the full model grid.
Comparing the trend of the fit results to the trend inherent in the
grid, it can be seen that the fit results represent values of mass,
age and temperature that are not biased by the trend in the model grid
but are probably those that selectively describe the observed data.

These results describe most of the sources as massive protostars with
a mean stellar mass of $\sim$10--20 \Msun and includes stellar masses
as high as 40 \Msun, with total luminosities between $10^3$ and
$10^5L_\odot$. Many of them are very young, with ages of
$\sim10^{3}-10^{4}$ years, and only a small fraction are modelled as
1Myr objects. It should be noted that the model grid becomes very low
in density after 5-10\Msun\, and even so, the observed SEDs have the
highest densities for the 10-30\Msun\, range.  The near distance
fitting naturally selects a relatively lower mass for the source
compared to the far distance fitting.  If HMPOs are in the accretion
phase, the estimated young ages are not a surprise because the
accretion phase is supposed to be very fast, typically 5$\times
10^{4}$yr or less \citep{mt03}

The model grid is roughly populated with constant density in log(age)
space (see fig.2 of \citet{rob06}). Curiously, the observed data
strongly picks up models that have lower ages with a peak between
$\sim10^{3.5}-10^{4.5}$ years. Luminosity peaks (not shown here) are
consistent with the observed IRAS luminosities of these sources.

The radii of the photospheres selected by the observed data are
typically large between 20-200\Rsun. The stellar temperature is found
to be typically between 4000K to 8000K but there is a considerable
number of models spanning a range between 10000K-26000K. It is
important to note that while the mass and age are uniformly sampled
for the grid limits, the radius and temperature are interpolated using
evolutionary models. Nevertheless the model grid covers a range of
stellar radii from a few solar radii to few hundreds of solar radii
and temperatures from 2500K to 40000K. It is interesting to note that
the observed SEDs preferentially select models with large stellar
radii(20--200Rsun) and relatively lower temperatures (4000K-8000K).
  
The SED is sensitive to the T$_{\star}$ and L$_{tot}$, which are
therefore basic results. The R$_{\star}$ and age are derived from
these parameters using evolutionary models.

\subsection{Disks and envelopes around massive protostars}

\begin{figure}
 \centering
 
       \includegraphics[width=9cm,angle=0]{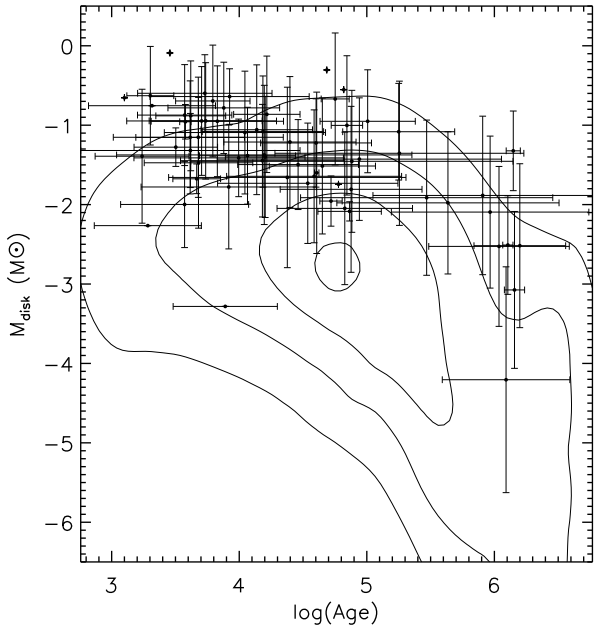}
  
       \caption{ Correlation between the disk mass and the source age
         estimated for all the sources with disks. Standard deviations for the two parameters are given by the error bars. If a given source was fitted by a single model, no error bars are shown. The contours represent the distribution of the same parameters in the entire grid of models used in this work. The outer contour corresponds to mean value and they increase in standard deviation units. }
 
 \label{fig:diskmass}
    \end{figure}

Fig.~\ref{fig:diskmass} compares the mass of the disk and the age of
the star with the associated spread in the values. The symbols
represent the weighted mean values and the error bars represent the
weighted standard deviations. The contours show the distribution
inherent to the model grid, with the outer most contour representing
mean values of the entire grid and subsequent contours are spaced at
intervals of 1$\sigma$ from the mean value. The data points from the
fitting results occupy a region representing higher disk masses
compared to the mean value of the grid. Thus, the modelled sources are
best represented with massive disks. It can be seen that for the
youngest sources, there is a spread of disk masses around $\sim$ 0.1
\Msun with no visible trend. However, for the few older sources at
$\sim10^6$ yrs, the disks become less massive and have larger
uncertainties, as one would expect.

\begin{figure*}
\centering

      \includegraphics[width=18cm,angle=0]{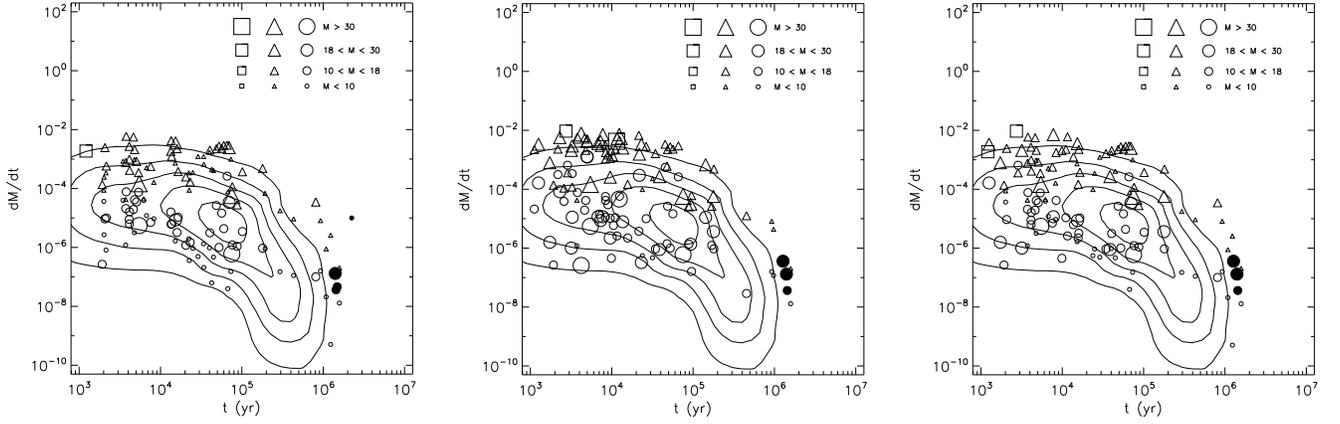}

      \caption{Correlations between the accretion rate and the age of
        the protostar. In the left panel we present the results with the near distances assumed for sources with distance ambiguity. In the middle panel, the same is shown with with far distances assumed. In the right panel, the parameters shown for these sources are for the distance with the lowest $\chi^2$ fit. The empty circles represent the disc accretion rate
        for all the sources with discs. The triangles represent the
        envelope accretion rate for all the sources with an envelope
        and a disc. Squares represent the envelope accretion rate for
        the sources without discs. Filled black circles represent
        sources without envelopes. The envelope accretion rate
        distribution inherent to the grid is shown by the contours as in fig.~6}
         
\label{fig:acctime}
   \end{figure*}

Fig.~\ref{fig:acctime} shows a plot of the accretion rate versus age
of the protostar for all the 68 fitted sources. The left, middle and
right panels represents results from near distance, far distance
fitting and the ``best $\chi^2$ distance'' respectively. Protostars
without disks are represented by squares, meaning they represent
envelope accretion rates alone.  For protostars with an envelope and a
disk together, empty circles correspond to the disc accretion rate and
the triangles correspond to the envelope accretion rate. Filled
circles represent disk accretion rates for sources without envelopes.
Naturally, there will be an empty circle associated with each
triangle. The sizes of all these symbols are shown roughly
proportional to the associated stellar mass which is distributed in to
four groups that are representative of O, B0-1, late B stars and A
type or lower mass stars. The Disk accretion rate is lower than
envelope accretion rate as expected from the assumed physics in the
models.  Protostars without envelopes (filled circles) have ages
greater than one million years.  The solid contours represent the
range of envelope accretion rates inherent to the model grid.

Most of the modelled sources show the presence of both disks and
envelopes when fitted with both near and far distance estimates. The
mean disk masses and accretion rates are $10^{-1.4\pm0.7}$ \Msun\, and
$\sim 10^{-6}$ \Msun yr$^{-1}$, respectively. The disk inner and
  outer radius range from a few AU to 50AU respectively. The dust
  sublimation radius is mostly within 10AU but can go as high 45AU
  depending on the mass and age of the modelled source. The envelope
accretion rates and sizes are $\sim 10^{-3}$ \Msun yr$^{-1}$ and 65000
AU respectively. In particular, the envelope sizes are large with a
spread between 50000 and 150000 AU, which corresponds to the limit at
which the model truncates the envelope radius. This distribution peaks
at $\sim$50000-60000 AU. At an average distance of 5kpc, it is very
similar to the projected sizes of the infrared nebulae found in Paper
I. The luminous star heats up large volumes that may not necessarily
be part of the infalling envelope, but rather the ambient molecular
cloud. The SED is sensitive to the stellar temperature and total
luminosity, from which the remaining parameters are derived. Therefore
the large envelope radii just represents the radius at which dust is
heated by the star.

    The dependence of the disk and envelope accretion rates as a
    function of stellar mass is seen as a power law. This power law
    dependence is estimated from the SED fitting results and compared
    against that which is inherent to the model grid in the same mass
    range. The envelope accretion rate from the SED fitting results
    scales as $\dot M_{env}$ = 10$^{-5.5\pm0.6} \times
    {M_*}^{2\pm0.6}$.  In contrast, the power law inherent to the
    model grid is $\dot M_{env}$ = 10$^{-5.3\pm0.02} \times
    {M_*}^{1\pm0.02}$ over the same mass range. This accretion rate
    dependence is indicative of spherical accretion \citep{bondi52}.
    Since the envelope sizes are linked to the luminosity, this
    relation simply means that the accretion is approximated to a
    spherical case at the scale of the dense core.  Similarly, the
    disk accretion from fitting results is related to the source mass
    as $\dot M_{disk}$ = 10$^{-7.2\pm0.7} \times {M_*}^{1.5\pm0.7}$
    and that inherent to model grid is $\dot M_{disk}$ =
    10$^{-9.7\pm0.03} \times {M_*}^{2.5\pm0.03}$.  Recently,
    \citet{fazal08} have modelled 13 similar targets from the Sri02
    lists using the same fitting tool and found very similar relation
    for the envelope accretion rates.  However, parameters estimated
    for individual sources by \citet{fazal08} can be different from
    the values listed in Table.1, owing to different data sets and
    selection criteria. In particular, they use MIPS observations (not
    used here) and also the integrated flux in the millimetre regime
    whereas we use the peak fluxes from millimetre data. Also, the
    relations given here are obtained from a larger sample,
    constraining the uncertainties in the power-law coefficients.

\subsection{Individual sources}

The following notes serve to identify the modelled sources in
conjunction with online figures from Paper I, or from the original
GLIMPSE images:

\textbf{13384--6152mms1} A high AM source matching well with the millimetre
peak. A clear point source with a unipolar nebula.

\textbf{13395--6153mms1} A highly saturated GLIMPSE point source
matching with the 1.2mm peak. Good photometry available in the JHK
bands.  GLIMPSE photometry not available due to saturation.

\textbf{13481--6124mms1} Highly saturated point source in GLIMPSE
image. JHK photometry available and no GLIMPSE photometry. All data
were used as points. The model fit indicates a 20 \Msun and young star
at $\sim$10$^4$ yrs.

\textbf{13560--6133mms1} An isolated high AM source coinciding with the mm
peak. The point source is surrounded by a shell like IR nebula. No
clustering or other YSO like sources within the shell.

\textbf{14000--6104mms3} A high AM point source in the middle of a large
diffuse nebula. GLIMPSE data for the point source is good. The mm data
was used as coming from a single source.

\textbf{14131--6126mms1} GLIMPSE point source surrounded by an extended
(scorpion shaped) 8 $\mu$m nebula. The 1.2mm data was used as a ''single source'' point. The point source had JHK data. IRAC ch1 and ch4 data not available probably because of contamination from the nebula.

\textbf{14166--6118mms1} Point source with photometry from GLIMPSE to mm.
It is surrounded by an IR nebula with a core-halo structure.

\textbf{14425--6023mms1} An isolated (nearly saturated) bright GLIMPSE
point source with H and K photometry as well. Given that there are no
other FIR/mm contributing nebula or stars in the field we used MSX,
IRAS and mm data as points.

\textbf{15246--5612mms1} A high AM point source surrounded by an IR nebula
which looks like a dense core irradiated by the central star. Data in
the H, K, GLIMPSE and the 1.2 mm bands were used as single source points.

\textbf{15347--5518mms1} A bright GLIMPSE point source surrounded by an IR
nebula with a core-halo shape. The 1.2mm data was assumed as a single source point.

\textbf{15506--5325mms1} Almost saturated GLIMPSE point source with a good
JHK counterpart. The mm emission and FIR fluxes do not represent the
fitted model that traces the JHK and MSX points. The IR point source
appears to be an evolved YSO. The IRAS and mm emission may be arising
from an adjacent, more deeply embedded object.

\textbf{15519--5430mms1} Well defined isolated GLIMPSE point source in the
middle of a large infrared nebula. The SED was well fitted by an
evolved 6 \Msun object. The IRAS and mm emission is not well
represented by the model.

\textbf{15579--5347mms1} A bright isolated point source surrounded by a
dense cocoon and compact nebula which extends out to an arm like
structure on the 8 $\mu$m image. The source appears in K band as well.

\textbf{16061--5048mms3} A chain of 6 point sources bordering an arc
shaped IR nebula on one side and an IR dark patch on the other side.
The sources seem to be inside the dark cloud that extends to a much
larger size.  The mm peak coincides with a fuzzy point source for
which photometry is available only in the ch3 and ch4. The IRAS and
MSX point are 30\arcsec off, therefore the mm data is used as a data point
and the IRAS as upper limits.

\textbf{16082--5031mms1} One bright star situated on the periphery of
a ring shaped nebula. The source also appears to be aligned with a
dark filamentary cloud on a larger scale. Although the source is seen
in GLIMPSE, the catalogs do not list the photometry. But a good 2MASS
counterpart exists. We used 2MASS and 1.2 mm data as points with IRAS
as upper limits.

\textbf{16093--5015mms2} Two bright high AM product sources adjacent to
each other. The mms2 source coincides well with one of the stars which
is also well centred on an IR dark cloud. A bright U shaped nebula is
adjacent to the source. The IRAS and MSX sources coincide with the
centre of this nebula.

\textbf{16148--5011mms2} A bright IR nebula inside which are embedded
two point sources. The brighter of the two coincides with mms2 and has
photometry in JHK, IRAC ch3 and ch4 bands. IRAC ch1 and ch2 photometry
are probably contaminated by the nebula. We used 1.2mm data as a point.

\textbf{16218--4931mms1} A clear point source with both 2MASS and
GLIMPSE data.  The source did not appear as high AM source in paper
I. The point source is linked to an IR nebula with a tail like
morphology. The fit was not very good.

\textbf{16219--4848mms1} An isolated IR source with 2MASS K, IRAC ch2 and
ch3 band photometry available, surrounded by an arc-shaped nebula. The
fit has a large degeneracy and the IRAS 100 $\mu$m emission is not
represented well by the fit.

\textbf{16232--4917mms1} A bright point source surrounded by a dense
cocoon and an arc shaped nebula with streamers connecting the arc and
the star. The star has photometry in 2MASS JHK and GLIMPSE and a shows
a well defined mm peak. MSX data was removed as it was not positionaly consistent with the other data and that was confirmed by the fitting results.

\textbf{16344--4605mms1} A filamentary IR dark cloud. At least eight
deeply embedded red sources are found to be coinciding clearly along
the filament. These sources are close to the mm peak. Based on
positional correlation with MSX and IRAS, the mms1 was chosen for
modelling. It has data from the GLIMPSE to 1.2mm bands. Although the
degeneracy of the fit was high, we did not use mm data to force a fit
because of multiple red GLIMPSE sources in the same field.

\textbf{16363--4645mms3/mms6} The region is filled with multiple red
sources. Two mm peaks namely mms3 and mms6 are associated with compact
GLIMPSE nebulae and point sources. They are also associated with two
separate MSX sources. The
mms6 source was modelled using the GLIMPSE and mm data as points. The
MSX was not used because the 8 $\mu$m image shows a bright compact IR
nebula nearby this source and could be contaminating the MSX flux.

\textbf{16464-4359mms1} Faint GLIMPSE point source with a small nebula.

\textbf{16501--4314mms1} Two IR point sources in the field, one at the
tip of a fan shaped nebula. This source coincides with mms1 and the
MSX peak and was modelled.

\textbf{16573--4214mms5} An isolated point source surrounded by a
bright IR nebula. No other confusing sources in the vicinity, and
a decent fit was obtained by using the 1.2 mm data as a point.

\textbf{18089-1732mms3} Three high AM sources surrounded by three fan
shaped infrared nebulae. One of this high AM sources coincides with
the mms3 clump from the 1.2mm MAMBO maps which is modelled. This
source is also studied using interferometer data (see Sec.5.4)

\textbf{18090--1832mms1} Data for this source is available from the near-IR to the mm wavelengths, however due to possible confusion with nearby objects, only the IRAC and 2MASS K bands were used as data points. 

\textbf{18144--1723mms1} A low source by the definition of Mol96. A single
bright IR source (slightly nebulous object) is found coinciding with
the mm peak. The sub-mm observations are obtained with SCUBA. The model is
constrained using GLIMPSE and MSX data.

\textbf{18151--1208mms1} No GLIMPSE point source data available for this
source. The source appears point like in the 2MASS K band and (sub)mm.
An extended nebulae/halo is found surrounding the point source in the K
band as well as GLIMPSE images. The source shows an outflow studied by \citet{dav04}.

\textbf{18159--1550mms1} Three point sources centrally surrounded by a bipolar
shaped nebula. Sub-mm source looks point-like. Millimetre image looks fuzzy
with no clear point source. (Not an high AM source).

\textbf{18247--1147mms1} A well isolated point source identified from the
NIR to (sub)mm peak.  The coordinates match to an accuracy of better
than  1\arcsec. We used the integrated fluxes in the (sub)mm because
of its point like appearance and treated everything as a point because
nothing else was found in a 20\arcsec radius.

\textbf{18264--1152mms1} Bright high AM product source in the middle of a
bright 8 $\mu$m nebula. The GLIMPSE source matches well with (sub)mm
peaks which appears point like despite the large infrared nebulae.

\textbf{18290--0924mms1} Apparently confusing object. A high AM source
situated 8\arcsec from the sub-mm peak. But the sub-mm peak has no
counterparts and shows up as a high contrast dark patch located
adjacent to the bright star and faint nebula. We modelled the high AM
source only using GLIMPSE, MSX+IRAS. Therefore, we assume the modelled point
source and the sub-mm emission as different entities.

\textbf{18306--0835mms2} The main (sub)mm peak coincides with a compact
nebula and does not have an IR point source. Roughly 1.5' away the
(sub)mm source 2 has a MSX point source and has a good IRC which is
modelled.

\textbf{18308--0841mms3} The main (sub)mm peak coincides with a bipolar IR
nebula that is at the tip of a dark filament. A fainter (sub)mm peak
(peak 3 of MAMBO) is associated with an IRC which is modelled using
GLIMPSE, MSX, MAMBO+SCUBA. There is no detected flux at 450 $\mu$m for
this source.

\textbf{18310--0825mms1} The (sub)mm peak is closely located to a GLIMPSE
source that is modelled. The MSX+IRAS peaks appear 15\arcsec offset
from the (sub)mm peak and GLIMPSE source and is coincident with an IR
nebula.

\textbf{18337--0743mms1} The (sub)mm peak coincides with a bright and
compact IR nebula in shape and position. The IR nebula is well
centred on a star which is visible in the JHK bands. The GLIMPSE
counterpart is not available (although visible on the images) in the
catalog. The nebula is a core-halo shaped object. The MSX+IRAS point
sources are offset by 15\arcsec falling on the edge of the nebula. The
model was constrained by using the 2MASS+sub-mm peaks as data points
while MSX, IRAS and MAMBO were used as upper limits. No other
suspicious objects in a vicinity of $>30$\arcsec.

\textbf{18345--0641mms1} The (sub)mm peak coincides well with two very
bright GLIMPSE sources that are close to saturation. A 2$\mu$m source is
coinciding. The (sub)mm, IRAS, MSX, GLIMPSE, 2MASS positions match to
better than 2\arcsec. Not found as a high AM source in paper I,
because the GLIMPSE source is not available in the highly reliable
catalog. 

\textbf{18372--0541mms1} A bright (sub)mm condensation with an extended
elliptical shaped halo. The high AM source is bright and situated
within 4\arcsec of the (sub)mm peak. Two red fainter GLIMPSE sources
are found in a radius of 6\arcsec\ from the high AM source. No 2MASS
counterpart. The IR nebula at 8 $\mu$m resembles a core-halo shape
with a halo of $\sim$10\arcsec. The (sub)mm peak flux is used as a
point because of the morphological resemblance over the whole range of
wavelength and no other significant objects are found within a
20\arcsec radius.

\textbf{18431--0312mms1} Elliptical shaped (sub)mm peaks at 850$\mu$m,
450$\mu$m and 1.2 mm. Core-halo IR nebula with a K band and
GLIMPSE star. 8 $\mu$m photometry not available (possibly because of the
nebular contamination). MSX source matches closely. No other
suspicious source within a radius of 10\arcsec. Therefore, the
(sub)mm peak fluxes were used as points for the fitting.

\textbf{18440--0148mms1} The sub-mm peak is point-like. The mm (MAMBO) peak
is extended and elliptically shaped. The IRC is a bright star with photometry available only in ch3 and ch4 bands. No other
confusing sources nearby within a 10\arcsec radius.

\textbf{18445--0222mms1} Two high AM sources with a bright IR nebula.
(sub)mm peaks are clearly point like. One of the high AM source
matches to better than 2\arcsec with the sub-mm peak which is
modelled. MSX not used. IRAS used as upper limits. GLIMPSE + sub-mm
used as points.

\textbf{18447--0229mms1} One high AM IRC matches well with the (sub)mm
peak. IRAS was used as upper limits while GLIMPSE and (sub)mm were
used as data point.

\textbf{18454--0136mms1/mms2/mms3} Three clean GLIMPSE point sources
with high AM product are within the FIR/sub-mm beams. We use the
GLIMPSE data as points and the FIR/mm data as upper limits for each of
the IR sources.  This is a cluster of at least 5 GLIMPSE sources
within the sub-mm beam and two of them are high AM sources. Three
bright sources are modelled. mms1 coincides with the mm peak. mms2 is
the brightest GLIMPSE source and has radio continuum emission.

\textbf{18460--0307mms1} A bright high AM source coincides with IRAS, MSX
and sub-mm peaks. We used only the IRAS 60 and 100 $\mu$m, K band,
(sub)mm and GLIMPSE points to get the fit. Some of the MSX and IRAS
photometry was removed to improve the fit.

\textbf{18472--0022mms1} One bright high AM source coincides with the
sub-mm peaks. The (sub)mm emission is extended but with a single well
defined peak. The IR source is surrounded by a cocoon in the IR and a
fan shaped nebula adjacent to it. The high AM source is modelled.

\textbf{18488+0000mms1} A bright star (saturated in GLIMPSE). This source
was not classified as high AM product in paper I because of
saturation. Photometry for ch1 and ch3 is available from GLIMPSE less
reliable catalog and a bright K band source coincides with the star.
The bright star is surrounded by an unipolar fan shaped nebula. The
(sub)mm emission is a circularly concentrated peak with a very faint
extension towards the unipolar flow. The ch3 photometry was treated as
upper limit (likely saturated) and 1.3 mm is treated as a point to
improve the $\chi^2$ of the fit.

\textbf{18511+0146mms1} A bright saturated GLIMPSE source coinciding
well with the mms1 millimetre core. The source did not appear as a
high AM product source in Paper I and no GLIMPSE photometry is
available due to its saturated nature. However, the star has good
2MASS J,H,K photometry. MSX data was used as points along with IRAS
and millimetre data as upper limits to obtain the fit.

\textbf{18521+0134mms1} One high AM source coinciding with
the (sub)mm peaks to better than 2\arcsec. The (sub)mm peaks are well
concentrated with a surrounding halo. The high AM source is modelled
with all data points.

\textbf{18530+0215mms1} There is no point source in the GLIMPSE or 2MASS.
GLIMPSE data shows a spectacular bipolar nebula typical of an outflow.
The (sub)mm emission is concentrated (point-like) with a faint
extension in the direction of BP outflows. We used the MSX, IRAS and
(sub)mm data, treated the (sub)mm data as points.

\textbf{18553+0414mms1} Two bright high AM sources close to each
other. The brightest of the two coincides with the (sub)mm peak. The
GLIMPSE ch4 data is saturated. There are no 2MASS counterparts. No
suspicious clustering objects nearby. The brighter source coinciding
with the (sub)mm peak is modelled.

\textbf{18566+0408mms1} This source is a very bright GLIMPSE source
(saturated in the ch3 and ch4) and only a K band counterpart is
available. The positional correlation with the
(sub)mm peak is very good ($<$2\arcsec). However, the source appears
as a double source, both comparable in brightness. This may be
indicative of the (sub)mm peak appearing elongated elliptically. This
is probably a case of massive binary. The source is modelled as a
single source.

\textbf{19012+0536mms1} The (sub)mm and FIR peaks appear point like.
GLIMPSE data reveals at least 5 point sources and a bright
concentrated IR nebula within the beams of (sub)mm and IRAS data.
The IR source coinciding with the (sub)mm peak is modelled . Indeed
this source appeared as a medium AM source in paper I and the high AM
source in the same field from paper I is 20\arcsec away from the
(sub)mm peak which is not modelled.

\textbf{19035+0641mms1} The sub-mm peak and all other data match to better
than 1\arcsec. The IR source appears bright in the GLIMPSE.

\textbf{19045+0813mms1} The IR source is visible in all Spitzer bands in
the 2MASS K band. The 1.2 mm data (peak flux) was obtained by
\citet{sanchez08}. Very good fit. The object in the IR is a bright
source with a surrounding infrared nebula that appears to show four arm
like structures. It can also represent wind blown cavity walls.

\textbf{19074+0752mms1} A high AM source aligned well with the (sub)mm
peak. An arc shaped unipolar nebula originating from the IR point
source. No photometry available for IRAC ch1 and ch2 bands.  By using
the Spitzer sources as points we produced a big discrepancy between the
best fit model and the remaining data used as upper
limits. We decided then to use the 450 $\mu$m data was used as a point.

\textbf{19092+0841mms1} A well defined GLIMPSE point source which
appeared as an average AM source in Paper I coincides well with the
millimetre peak. The point source is surrounded by a fan shaped
unipolar infrared nebula and also visible in the 2MASS K band. The K
band and GLIMPSE photometry were treated as data points with all the
other as upper limits to obtain a fit. This source was modelled with a
high degeneracy.

\textbf{19175+1357mms1} The (sub)mm peak coincides with a compact, bright
8 $\mu$m nebula in which there appears to be embedded 3 point-like
sources. However, the catalog lists only one point source for which
data is available only in ch3. The photometry in other bands is
contaminated or was difficult to obtain because of the bright nebula.
Since at least 3 points are needed to the RT fit and the (sub)mm
emission is concentrated like a point source or a peak, we use 450 and
850 $\mu$m data as points. 

\textbf{19217+1651mms1} One of the best studied and clean sources
using photometry all the way from 2MASS K band to the interferometric
observations.  No contamination from any adjacent sources. 

\textbf{19266+1745mms1} A GLIMPSE point source with an IR nebula. The
(sub)mm data shows emission that looks like a tadpole with a peak and
a tail (extending to the south). 

\textbf{19282+1814mms1} The (sub)mm source appears extended with one
prominent peak where the MSX and GLIMPSE sources coincide. The IRAS
source is about 60\arcsec off to the east (we did not use it). The
region is filled with several GLIMPSE point sources, with at
least 10 red sources surrounding the modelled bright source.
In paper I, this bright source did not appear as a high AM source.
Submm data was used as a data point.

\textbf{19410+2336mms1} A satured GLIMPSE source represent the (sub)mm
and FIR peaks to better than 1\arcsec. Data for this was found in the
less reliable catalog. However a good K band data point constrains the
fit. So we used 450/850 $\mu$m and K band as points and removed the
GLIMPSE data. A well defined outflow originates from this mm source.

\textbf{19410+2336mms2} A faint GLIMPSE source without MSX and IRAS
counterparts matches the sub-mm peak. A well defined outflow is found
to originate from this mm source.

\textbf{19411+2306mms1} The IR source appears in the 2MASS JHK and the
GLIMPSE data as a point source but shows extended (sub)mm
emission. No other high AM product source in the vicinity. The (sub)mm, MSX, 2MASS
and GLIMPSE were all used as data points. This model is
highly constrained and defined by the short wavelength data.

\textbf{19413+2332mms1} The GLIMPSE images display two fan shaped
nebulae separated by a dark lane. A faint GLIMPSE source which has
photometry in ch1 and ch2 and also in the 2MASS K band, is located
coinciding with the tip of the fan shaped nebula and also with the
millimetre core. The sub-mm data appears point like and the SCUBA data was used as a point.

\section{Discussion}

\subsection{Caveats inherent to the modelled data}

In the previous section, the physical parameters of the star, disk and
envelopes were estimated by fitting the SEDs of the IR point sources.
The resulting variables were compared to identify specific trends. In
the following we will discuss some caveats pertinent to the
observational data, fitting procedure and selected sample.

\begin{itemize}

\item{Masses of the driving engines}

  The SED fitting results show that some objects with the highest
  stellar masses (luminosities) are also among the youngest sources in
  the sample. Older sources with similar masses are expected to be
  already in an evolved HII region stage and are probably not detected
  because their FIR colours would not be in agreement with the initial
  criteria of the selection of HMPO candidates and/or they would have
  significant ionised emission \citep[See e.g.][]{kk08}. On the other
  hand, the less massive sources in this sample are concentrated
  towards older ages. At a distance of a few kpc, the GLIMPSE data
  would be sensitive to detect pre-main sequence objects rather than
  embedded protostars of lower masses. Besides this, the most massive
  protostars (early O stars) may be residing in the infrared-dark
  clouds (IRDC's) \citep{jack08} that shine only in the FIR and
  millimetre wavelengths or associated with the driving engines of HII
  regions.

  The models of \citet{rob06} do not account for the accretion
  luminosity arising in the envelopes which is thought to be
  significant for sources with masses greater than 20\Msun. However,
  they do compute the disk accretion luminosities for all mass ranges.
  The results from this work and in Paper I, indicates that envelope
  accretion or accretion through gaseous disks may play an important
  role in building the most massive stars.  Therefore the few sources
  modelled with M$\ge$20\Msun may need alternative investigation.
  Naively, this means that stellar masses $>$20\Msun are probably
  over-estimates and the sample studied here represents a good census
  of proto-B stars rather than proto-O stars. However, some of these
  proto-B stars at present time could become O stars in future.

\item{Multiplicity issue}

  One of the important aspects of massive star formation is that they
  are known to form in multiple systems or strongly clustered
  environments. In Paper I we discussed the issues related to
  clustering around the IRCs of HMPOs analysed here and concluded that
  we did not find clustering at the level of $\sim$4-5\Msun\ members.
  The ``isolated'' nature is therefore valid only for the mass range
  beyond 10\Msun\ . These objects can, however, have multiplicity at a
  level of $\sim$6000AU (at a distance of 3\,kpc) owing to the assumed
  2\arcsec spatial resolution of the GLIMPSE data. Assuming a source
  in the simplest case of multiplicity - a binary system, there are
  two possible scenarios; a) two objects at the same evolutionary
  state and b) objects in different evolutionary states. In the first
  case, the SED modelling can not detect possible multiplicity because
  the SED shapes will be similar for the two sources and may overlap.
  In the latter case, where sources are at different evolutionary
  stages, the SEDs are expected to flatten out because of the older
  source and is prone to be detected by its unfamiliar shape. It turns
  out that we did not find any source from our fitting which deviates
  from the normal single source SED with a single exception
  (18447-0229mms1), and therefore multiplicity of sources with
  different evolutionary stages is not a likely possibility. See the
  online Fig.1 to check this.

\end{itemize}

\subsection{Ionised accretion flows?}

As mentioned previously in Paper I, free-free continuum emission in
the centimetre regime coinciding with 27 HMPOs in that list was
detected with VLA and other observations.  Free-free continuum
emission indicates the presence of compact ionised regions surrounding
the massive protostars and/or winds in the bipolar cavities. Eleven of
the 27 sources were modelled in this work and are shown as filled grey
symbols in Fig.~8. The fitted models describe massive stars with an
actively accreting disk and envelope. But their properties such as
accretion rates, ages, or masses are not different from the mean
values for the rest of the sample.

Therefore these sources could be typical YSOs with ongoing accretion
and significant ionised gas close to the star, making them good
examples for accretion past the HII region.  If the accretion scenario
suggested by the models is in fact correct, the presence of ionised
emission could imply the existence of ionised accretion flows and/or
ionised winds, in addition to the molecular components.  The scenario
of continuing accretion (accretion past the HII region) therefore
could play an important role in the most massive star formation.

Although several massive protostars are found from this work, some of
the most massive objects from the initial sample of HMPO candidates
are still missing from this study. This is because such sources appear
very bright in the near/mid IR bands and are saturated in the IRAC
images (see Paper I). Such sources can make a good target sample to
investigate the mechanism of accretion past the HII region, by using
short exposure ground based IR and centimetre continuum observations.

\subsection{HR Diagram}

\begin{figure}

      \includegraphics[width=9cm,angle=0]{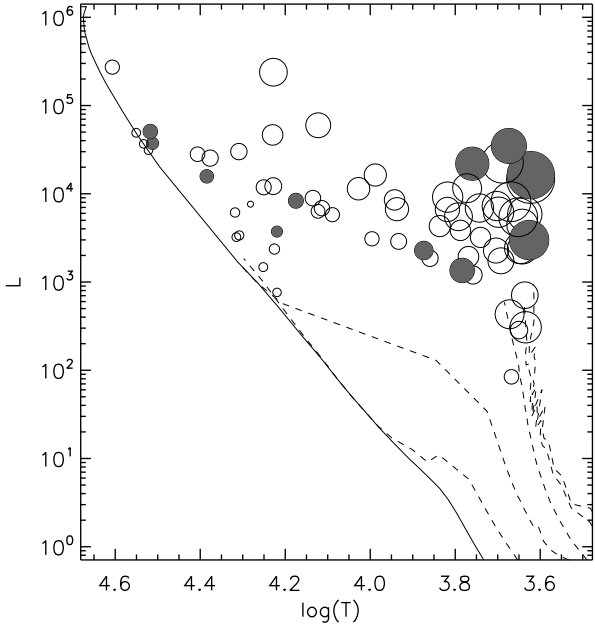}
     
      \caption{Luminosity vs. stellar temperature plot. Filled circles
        represent sources with known radio continuum emission as found
        in the literature.
        Solid line marks the birth-line track from \citet{bm96} for masses
        above 7\Msun and represent the ZAMS up to 7\Msun. The dotted
        lines are Geneva isochrones up to 7\Msun. From left to right
        the dotted lines represent values of log(age)= 6,5,4 and 3.}

         \label{fig:lm}
   \end{figure}

   In Fig.~\ref{fig:lm} the luminosity and stellar temperature of all
   the modelled sources are compared to simulate an HR diagram. It is
   important to note in advance, that this HR diagram is prone to
   mimic the assumed stellar evolutionary models. Nevertheless, some
   facts are worth noting. The shaded circles represent the sources
   associated with significant free-free continuum emission, but the
   contrary is not true. There is simply no radio observations
   available for all the sources to ascertain the fact. The size of
   the symbols are made inversely proportional to the age of the
   protostar in this figure. Clearly, the oldest sources (smallest
   circles) trace the zero-age main sequence locus. These sources have
   smaller accretion rates and the stellar photosphere is a
   significant contributor to the total luminosity. The youngest
   sources (big circles) are all found above the 'main sequence'
   locus. At younger ages, other physical processes such as envelope
   accretion are the dominant contributors to the total luminosity and
   the photosphere is mostly obscured.  Sources found occupying the
   right side of the diagram are likely protostars representing the
   onset of deuterium burning along with the presence of a convective
   envelope. The observed shapes and luminosities of the SED clearly
   exclude the possibility of classical intermediate PMS stars (dotted
   lines are PMS tracks up to 7\Msun ). The sources associated with
   radio continuum emission are distributed in different parts of the
   HR diagram and do not show any particular trend with other
   estimated properties. Four sources, found on the extreme right hand
   side of the diagram are worth noting for their youth, mass and
   association with radio continuum emission. They may represent
   sources where powerful ionised winds or accretion flows are onset
   at a very early stage of formation. But it should also be noted
   that the age estimates from the SED fitting are probably uncertain.
   The difference between accreting and non-accreting massive star on
   the main sequence could be identified better by observing the
   structure of the ionised region at smaller scales
   \citep{mt03,kk08}.

\subsection{How real are the driving engines, disks and envelopes?}

Throughout this paper, we focussed on how the observed data
preferentially chooses certain set of values or relations from a
larger set of available options in the model grid. While these results
may appear real, it is important to view them within the limitations
of the models, discussed in \citet{rob06} and remembering ``what you
put in is what you get out'' quoting \citet{rob08}.

The SED modelling selectively picked up large radii for driving
engines and lower temperatures. Both these parameters are derived from
the assumed evolutionary tracks after uniformly sampling the mass and
age grid in the models. The accretion rates assumed in computing those
evolutionary tracks are typically $10^{-4}$--$10^{-5}$ \Msun
yr$^{-1}$.  The SED fitting results also suggest that the accretion
rates are much higher than those values and could be as high as
10$^{-2}$--10$^{-3}$\Msun yr$^{-1}$. In a situation like that, the
mass- radius relationships for the massive protostars are also
expected to be different. \citet{hos08} discuss such effects in
detail. In any case, although the SED modelling suggest photospheres
with certain properties, it is far from clear if those are real
photospheres where hydrogen burning has begun or if those surfaces are
totally different type of entities \citep{mt03,zy07}.

The SEDs for high mass protostars are dominated by the envelope flux
longward of 8-10$\mu$m and the disk flux is almost always embedded
within the envelope flux. It is possible, therefore, that models with
envelopes alone and models with disks and envelopes successfully fit
the same SED. It is particularly true in the case of sources without
photometry in the near-IR or IRAC ch1. Although, the results here
favour the presence of massive disks with relatively larger radii, the
overall SED may not be sensitive to the presence or absence of the
disk. But the envelope geometry is indirectly dependent on the disk
radius, since the models use a rotating-infalling solution. 
  Therefore, in the few cases, when the SED models indicates no disk,
  it may simply mean that the disk is gaseous and its size is well
  within the dust destruction radius which could be between a few AU
  to 50AU for the sample studied here. \citet{bik04} argued for the
  presence of disks in massive young stars (much older than the ones
  in this sample) by modelling the CO band heads with Keplerian disks.
  They estimated small inner and outer disk radii (0.2--3.6AU) and
  masses ($<$30\Msun) for those presumed molecular disks.
  \citet{grave07} conducted a spectro-astrometric analysis of the
  \citet{bik04} data and found that the disk sizes should be much
  larger (200-300AU) than originally predicted by those authors. In
  the light of SED fitting results, the \citet{bik04} results conform
  with the pure gaseous disks and those of \citet{grave07} may
  represent the gas+dust disks indicative of the rotating-infalling
  solution centrifugal radii.

We noted in Sec.~4.3 that the envelope sizes selected by the observed
data are preferentially large and that the size distribution is very
similar to the size distribution of IR nebulae found in Paper I. It is
worth recalling the definition of an envelope from \citet{rob06}. It
is defined as the distance at which the optically thin radiative
temperature falls to 30K. Therefore, it is no surprise that the sizes
of the modelled envelopes are similar to the sizes of the 8 $\mu$m
nebulae from the GLIMPSE images. These envelopes, therefore, are not
necessarily clean geometrical structures that extend to 0.3-0.5pc. It
is just the dimension at which ambient cloud is heated up due to the
luminosity of the central source.  High resolution interferometric
observations of massive protostellar candidates largely display
toroids with non-keplerian velocity profiles. These toroids are found
to have sizes of 10000-20000 AU and have some clean geometrical
structures \citep{bel05, fur08}. Although the present model grid takes
into account these factors, better observational constraints on the
toroid sizes, cavity opening angles and densities at larger distances
will be of important value to future model grids.

In summary, the observed SEDs are well explained in the frame work of
an accretion scenario model, but the true existence or nature of the
protostellar components around a massive protostar are still
debatable.

\section{Conclusions}

The near-infrared to millimetre spectral energy distributions of a
sample of 68 bonafide high mass protostellar objects were constructed
using the 2MASS, the version 2.0 catalogs of GLIMPSE, MSX, IRAS and
(sub)mm surveys. The resulting SEDs were modelled using a grid of
radiative transfer models and the following analysis allows us to
conclude the following:

\begin{itemize}

\item{ The observed SEDs of all the targets are best described by
    models of YSOs with a stellar mass in the range 5--40 \Msun, with
    a median mass of 12\Msun and with ages between 10$^3$--10$^6$ yrs
    with a median age of 10$^4$yr's. The observed data selects models
    with larger stellar radii of 20-200\Rsun and lower temperatures of
    4000-10000K.}

\item{The disk and envelope accretion rates estimated from SED
    modelling are concentrated around $\sim$10$^{-6}$\Msun /yr and
    $\sim$10$^{-3}$\Msun /yr respectively. The envelope accretion
    scales as a power law of stellar mass $\dot M_{env}$ =
    10$^{-5.5\pm0.6} \times {M_*}^{2\pm0.6}$, which added to the large
    envelope size merely implies that the accretion is spherical on
    the size-scales of dense cores.  Although the models show disks
    for each protostar, the SED may not be sensitive to the presence
    or absence of a disk, because the envelope flux almost always over
    powers the disk flux.  Studies of disks will need alternative
      investigations, nevertheless, it is indirectly implied here by
      the envelope solutions which have non-zero centrifugal radii.}

\item{ For the sample studied here, we exclude the possibility of
    multiplicity involving sources of different evolutionary states
    since the observed SEDs do not have a flattened appearance
    expected from such a ensemble. Even if multiplicity exists, the
    SED is dominated by the most luminous source, thus the stellar
    parameters are reasonable estimates.}

\item{ A fraction of sources modelled with an actively accreting disk
    and envelope also show centimetre free-free emission and known HII
    regions. Therefore, the mechanism of accretion past the HII region
    with ionised accretion flows may be important in building the most
    massive stars.}

\end{itemize}

\begin{acknowledgements}
  Our sincere thanks to the referee Barbara Whitney for her insightful
  review that greatly improved the paper and also taught us a
  ``matured usage'' of the SED fitting tool . Grave and Kumar are
  supported by a research grant PTDC/CTE-AST/65971/2006 approved by
  the FCT (The Portuguese national science foundation).  Grave is also
  supported by a doctoral fellowship SFRH/BD/21624/2005 approved by
  FCT and POCTI, with funds from the European community programme
  FEDER.

\end{acknowledgements}

\bibliographystyle{aa}

\end{document}